\documentclass{article}
\usepackage{cite}
\usepackage{graphicx}
\usepackage{dcolumn}

\begin{document}

\title{On the construction of pseudo-Hermitian Hamiltonians by means of similarity
transformations}
\author{Francisco M. Fern\'{a}ndez \thanks{%
E-mail: fernande@quimica.unlp.edu.ar} \\
INIFTA (UNLP, CCT La Plata-CONICET), Divisi\'on Qu\'imica Te\'orica\\
Blvd. 113 S/N, Sucursal 4, Casilla de Correo 16, 1900 La Plata, Argentina}
\maketitle

\begin{abstract}
We generalize a recently proposed approach for the construction of
pseudo-Hermitian Hamiltonians with real spectra. Present technique is based
on a simple and straightforward similarity transformation of the coordinate
and momentum.
\end{abstract}

\section{Introduction}

In a recent paper Miao and Xu\cite{MX16} derived two pseudo-Hermitian
Hamiltonians that have real spectra because they are isospectral to
Hermitian Hamiltonians. The approach is based on the Heisenberg equations of
motion for the coordinate and momentum. The similarity transformation
between the Hermitian and pseudo-Hermitian Hamiltonians is expressed in
terms of an infinite series of either the coordinate (first model) or
momentum (second model).

The purpose of this short article is to show that one can easily derive
those pseudo-Hermitian Hamiltonians by means of a considerably simpler and
more straightforward technique developed recently\cite{F16}.

\section{The similarity transformation}

Let $x$ and $p$ be the dimensionless coordinate and momentum, respectively,
that satisfy the commutation relation $[x,p]=i$. If $f(x)$ is a
differentiable real function of $x$ we have
\begin{equation}
e^{f}p^{2}e^{-f}=\left( p+if^{\prime }\right) ^{2}=p^{2}-f^{\prime
2}+i\left( f^{\prime }p+pf^{\prime }\right) ,
\end{equation}
where the prime denotes differentiation with respect to $x$. Therefore
\begin{equation}
e^{f}\left( p^{2}+V+f^{\prime 2}\right) e^{-f}=p^{2}+V+i\left( f^{\prime
}p+pf^{\prime }\right) ,  \label{eq:Similarity_1}
\end{equation}
where $V=V(x)$, is an arbitrary function of $x$. If we choose $V(x)$ and $%
f(x)$ so that $H_{H}=p^{2}+V(x)+f^{\prime }(x)^{2}$ supports real
eigenvalues then the non-Hermitian Hamiltonian $H=p^{2}+V(x)+i\left[
f^{\prime }(x)p+pf^{\prime }(x)\right] $ will exhibit exactly the same real
eigenvalues\cite{F16}.

If we now take into account that
\begin{equation}
e^{-f}p^{2}e^{f}=\left( p-if^{\prime }\right) ^{2}=p^{2}-f^{\prime
2}-i\left( f^{\prime }p+pf^{\prime }\right) ,
\end{equation}
then it follows from equation (\ref{eq:Similarity_1}) that
\begin{equation}
e^{-f}\left( p^{2}+V+f^{\prime 2}\right) e^{f}=p^{2}+V-i\left( f^{\prime
}p+pf^{\prime }\right) =e^{-2f}\left[ p^{2}+V+i\left( f^{\prime
}p+pf^{\prime }\right) \right] e^{2f}.  \label{eq:pseudo_Hemitian_1}
\end{equation}
In other words, $H$ is $\eta $-pseudo-Hermitian\cite{M02a,M02b,M02c}
\begin{equation}
H^{\dagger }=\eta ^{-1}H\eta ,
\end{equation}
where$\;\eta =e^{2f}$.

If $\psi $ is an eigenfunction of $H_{H}$ with eigenvalue $E$, then $%
e^{f}\psi $ is eigenfunction of $H$ with the same eigenvalue and we should
choose $V(x)$ in such a way that $e^{f}\psi $ is square integrable in the
case of a bound state.

The technique proposed by Miao and Xu\cite{MX16} is a particular case of the
one developed above when
\begin{equation}
f(x)=-\sum_{k=0}^{\infty }\frac{c_{k}}{k+n+1}x^{k+n+1},\;n\geq 0.
\label{eq:f(x)_MX}
\end{equation}
Present approach is somewhat more general because the function $f(x)$ does
not necessarily have to be of this rather restricted form. For example, if
we choose
\begin{equation}
V(x)=\frac{3D}{4}\left( 1-e^{-\alpha x}\right) ^{2},\;D>0,\;\alpha >0,
\end{equation}
and
\begin{equation}
f(x)=\frac{\sqrt{D}}{2}\left( x+\alpha ^{-1}e^{-\alpha x}\right) ,
\end{equation}
then we obtain a pseudo-Hermitian operator with the spectrum of the Morse
oscillator\cite{EWK44}
\begin{equation}
H_{H}=p^{2}+D\left( 1-e^{-\alpha x}\right) ^{2}.
\end{equation}
Note that this model exhibits discrete spectrum for $0<E<D$ and continuous
one for $E>D$.

In order to derive the second model we simply reverse the roles of the
coordinate and momentum and write the similarity transformation in terms of
a function $g(p)$:
\begin{equation}
e^{g}x^{2}e^{-g}=\left( x-ig^{\prime }\right) ^{2}
\end{equation}
In this case we should choose a suitable function of the momentum $V(p)$ in
order to obtain an Hermitian Hamiltonian with real spectrum\cite{MX16}.

It is quite easy to generalize the results above to a
quantum-mechanical system described by a set of $N$ coordinates
$\mathbf{x}=(x_{1},x_{2},\ldots
,x_{N})$ and their conjugate momenta $\mathbf{p}=(p_{1},p_{2},\ldots ,p_{N})$%
. In this case we obtain the isospectral Hamiltonians
\begin{eqnarray}
H_{H} &=&\mathbf{p}\cdot \mathbf{p}+V(\mathbf{x})+\nabla
f(\mathbf{x})\cdot
\nabla f(\mathbf{x})  \nonumber \\
H &=&\mathbf{p}\cdot \mathbf{p}+V(\mathbf{x})+i\left[ \nabla f(\mathbf{x}%
)\cdot \mathbf{p}+\mathbf{p}\cdot \nabla f(\mathbf{x})\right] .
\end{eqnarray}

\section{Conclusion}

In this short article we have shown that an approach for the construction of
pseudo-Hermitian Hamiltonians with real spectra developed recently\cite{F16}
is simpler and more straightforward than the one proposed by Miao and Xu\cite
{MX16} in two main aspects. First, our technique does not require resorting
to the Heisenberg equations of motion; second, it avoids the use of a
particular power series for the function that appears in the similarity
transformation. In this way, present method facilitates the construction of
pseudo-Hermitian Hamiltonians that are isospectral to Hermitian ones.


\begin{thebibliography}{9}
\bibitem{MX16}  Y.-G. Miao and Z.-M. Xu, Investigation of non-Hermitian
Hamiltonians in the Heisenberg picture, Phys. Lett. A 380 (2016) 1805-1810.
arXiv:1212.6705v5 [quant-ph].

\bibitem{F16}  F. M. Fern\'{a}ndez, Non-Hermitian Hamiltonians and
Similarity Transformations, Int. J. Theor. Phys. 55 (2016) 843-850.
arXiv:1502.02694v2 [quant-ph].

\bibitem{M02a}  A. Mostafazadeh, Pseudo-Hermiticity versus PT symmetry: The
necessary condition for the reality of the spectrum of a non-Hermitian
Hamiltonian, J. Math. Phys. 43 (2002) 205-214.

\bibitem{M02b}  A. Mostafazadeh, Pseudo-Hermiticity versus PT-symmetry. II.
A complete characterization of non-Hermitian Hamiltonians with a real
spectrum, J. Math. Phys. 43 (2002) 2814-2816.

\bibitem{M02c}  A. Mostafazadeh, Pseudo-Hermiticity versus PT-symmetry III:
Equivalence of pseudo-Hermiticity and the presence of antilinear symmetries,
J. Math. Phys. 43 (2002) 3944-3951.

\bibitem{EWK44}  H. Eyring, J. Walter, and G. E. Kimball, Quantum Chemistry,
John Wiley \& Sons, New York, 1944)
\end{thebibliography}
\end{document}